\documentclass[english]{lni}
\usepackage{algorithmic}
\usepackage{graphicx}
\usepackage{textcomp}
\usepackage{xcolor}
\def\BibTeX{{\rm B\kern-.05em{\sc i\kern-.025em b}\kern-.08em
    T\kern-.1667em\lower.7ex\hbox{E}\kern-.125emX}}

\usepackage{xspace}
\usepackage{subcaption}

\newcommand{\feedForfood}{Feed4Food\xspace} 
\newcommand{\european}{European\xspace} 
\newcommand{\bucharest}{Bucharest\xspace} 
\newcommand{\drama}{Drama\xspace} 
\newcommand{\strovolos}{Strovolos\xspace} 
\newcommand{\Roma}{Roma\xspace} 
\newcommand{\Romania}{Romania\xspace} 
\newcommand{\FallCommunism}{the fall of communism\xspace} 
\newcommand{\Arab}{Arab\xspace} 
\newcommand{\Moldovans}{Moldovans\xspace} 
\newcommand{\Greece}{Greece\xspace} 
\newcommand{\Cyprus}{Cyprus\xspace} 
\newcommand{\SteliosChristosIoannou}{Stelios Christos Ioannou\xspace} 
\newcommand{\Africans}{Africans\xspace} 
\newcommand{\Syrians}{Syrians\xspace} 
\newcommand{\Asians}{Asians\xspace} 

\newcommand{\LL}{LL\xspace} 
\newcommand{\LLs}{LLs\xspace} 

\newcommand{\etc}{\textit{etc. }}
\begin{document}
\title[]{Supporting a Sustainable and Inclusive Urban Agriculture Federation using Dashboarding}
 \author[1]{Klervie Toczé}{k.m.tocze@vu.nl}{0000-0002-7300-3603}
 \author[1]{Iffat Fatima}{i.fatima@vu.nl}{0000-0002-2430-0441}
 \author[1]{Patricia Lago}{p.lago@vu.nl}{0000-0002-2234-0845}
 \author[1]{Lia van Wesenbeeck}{c.f.a.van.wesenbeeck@vu.nl}{0000-0002-6710-7596}%
 \affil[1]{Vrije Universiteit Amsterdam\\De Boelelaan 1105\\1081 HV Amsterdam\\The Netherlands}
\maketitle

\begin{abstract}

Reliable access to food is a basic requirement in any sustainable society. However, achieving food security for all is still a challenge, especially for poor populations in urban environments. The project \feedForfood aims to use a federation of Living Labs of urban agriculture in different countries as a way to increase urban food security for vulnerable populations.

Since different Living Labs have different characteristics and ways of working, the vision is that the knowledge obtained in individual Living Labs can be leveraged at the federation level through federated learning. With this specific goal in mind, a dashboarding tool is being established.

In this work, we present a reusable process for establishing a dashboard that supports local awareness and decision making, as well as federated learning. The focus is on the first steps of this creation, \ie defining what data to collect (through the creation of Key Performance Indicators) and how to visualize it. We exemplify the proposed process with the \feedForfood project and report on 
our insights so far. 
\end{abstract}
\section{Introduction}
\label{sec:Introduction}

Securing food access for everyone is a basic prerequisite for a sustainable society and part of the second Sustainable Development Goal of the United Nations~\cite{unitednations17GOALSSustainable}. Recent crises worldwide (\eg extreme weather events, the COVID-19 pandemic, ongoing wars) have highlighted that food security is easily at risk~\cite{BlayPalmer_CityRegionFoodSystems,carducci2021food,kovacs2022war}.  In particular, vulnerable communities (\eg low-income consumers and consumers for whom access to food is difficult because of physical or mental challenges or social exclusion) living in urban environments experience difficulties in securing food security in case of a crisis~\cite{battersby2018linking,GonzalezUrbanAgricultureHavana}. Therefore, inclusive and eco-friendly urban agriculture represents a promising way to improve the food security of vulnerable urban populations and contribute to a sustainable society. 
Urban agriculture can take various forms, from cultivation on balconies to peri-urban farms.

The \feedForfood project 
uses pilot Living Labs (LLs) spread over three different \european cities to showcase how sustainable urban agriculture can empower and include vulnerable groups. The project will also promote the transition towards low-impact and regenerative urban food systems. Therefore, on the scale of each \LL, the vision of the project is to create awareness about the current status and challenges with regards to sustainability. On a larger scale, it is to realize a federation of \LLs and achieve federated learning, where the \LLs can learn from the failures and successes of all the other \LLs part of the federation. To achieve this vision, supporting (digital) tools are needed to monitor the \LLs' progress towards their objectives and to enable the knowledge exchange within the federation. 

Using dashboards is an established practice in companies to monitor business-related indicators in order to improve company goals, such as profitability. Recently, dashboards are increasingly used by a wider range of organizations, also coming from the public sector~\cite{maheshwari_dashboards_2014}. The type of monitored indicators is also widening, and sustainability-related indicators, in particular carbon emissions or energy usage, are increasingly being monitored, due to increasing regulatory pressure on ESG reporting~\cite{csrd} (\eg by the European Union). 
From their initial purpose of focusing on monitoring and decision support, dashboards now also focus on supporting learning and communication~\cite{sarikaya_what_2019,pauwels_dashboards_2009}. 
Therefore, a dashboard appears as the appropriate tool to support the \LLs federation in monitoring their progress, taking informed decisions,  creating awareness and achieving federated learning. 
  
A \feedForfood dashboard should be tailored to the needs and objectives of the urban \LLs that the tool will support. Hence, this work studies the following research question (RQ): 
How to establish a dashboard that supports a sustainable and inclusive urban agriculture federation?
This establishment poses two important challenges. First, the right data should be collected, \ie all (and only) the data relevant for the decisions to be taken (Challenge 1). Second, the right information should be provided to the dashboard user so that they can understand/make sense of the presented data to make informed decisions (Challenge 2).
The contributions of this work include:
\begin{itemize}
    \item A reusable process for establishing a dashboard to support a federation of user groups.
    \item The application of this process to the \feedForfood project, including the key performance indicators (KPIs) created.
    \item Our insights obtained when applying the process, and the challenges we identified. 
\end{itemize}

The rest of this article is structured as follows: related works are presented in Section \ref{sec:RelatedWorks} and the \feedForfood project and its vision are introduced in Section \ref{sec:Project}. Then, the proposed dashboard establishment process is introduced in Section \ref{sec:ProcessDescription}. This process has two parts: the 
KPI Creation process (Section \ref{sec:ViewKPI}) and the Dashboard Design process (Section \ref{sec:ViewArchitecture}). 
Insights obtained and challenges identified are described in Section \ref{sec:Discussion}. Section \ref{sec:Conclusion} concludes this article.

\section{Related Works}
\label{sec:RelatedWorks}

The term \textit{dashboard} encompasses a very diverse set of practices~\cite{sarikaya_what_2019}. The traditional definition, as given by Few~\cite{few2006information}, refers to ``a visual display of the most important information needed to achieve one or more objectives, consolidated and organized on a single screen so that the information can be monitored at a glance''. However, the concept of dashboard has evolved from the focus on showing all information in a single view to include multiple views with interactive interfaces~\cite{sarikaya_what_2019}. Therefore, new broader definitions such as the one by Wexler have appeared: ``a visual display of data used to monitor conditions and/or facilitate understanding''~\cite{wexler_big_2017}. The purpose of dashboards has also evolved, moving from the traditional monitoring of performance and decision support to cover, \eg, communication, learning, and forecasting~\cite{sarikaya_what_2019,pauwels_dashboards_2009}. 
This type of tool is often used by companies and increasingly by public sector organizations~\cite{maheshwari_dashboards_2014}. One of the first development stage of a dashboard is the selection of key indicators to be monitored~\cite{pauwels_dashboards_2009}.
A KPI is \textit{``an indicator that focuses on the aspects of organizational performance that are the most critical to the current and future success of the organization''}~\cite {parmenter2015}. These KPIs help decision-makers make informed decisions for the achievement of their goals. 

Some efforts have been made to develop KPIs and dashboards for (urban) agriculture. 
Wolfert \etal~\cite{WOLFERT2022} use the SDGs to develop KPIs for the sustainability monitoring of Internet of Things-based interventions in five agriculture subsectors. These sectors pose challenges like difficulty in setting appropriate target thresholds due to the lack of baseline data and unidentified end users, leading to evolving target values over time. The reliability of the data and the balance between sustainability and business needs is also challenging.

In the dairy sector, Freitas \etal~\cite{FREITAS2025} present a benchmarking platform using zootechnical and economic KPIs that support decision-making for dairy farm stakeholders. The KPI values are compared against benchmarks based on user-defined or pre-defined cohorts based on farm attributes. The dashboard also makes recommendations for farm improvements based on data-driven algorithms. 
The platform is validated using historical data from a large number of farms and market data. 

Vannieuwenhuyze~\cite{Vannieuwenhuyze2020} presents a dashboard for policy evaluation and decision-making about street gardens in Antwerp. The study emphasizes that the choice of KPIs is critical, as they drive the data collection process. The study presents a financial profit KPI with the concept of subjective interpretation of KPIs by allowing the dashboard users to change the model parameters based on their personal beliefs and assumptions. Due to the cyclic changing nature of the policy, the dashboard is proposed to include context information for better interpretation of the data. 

In recent years, increasing regulation regarding ESG reporting~\cite{csrd} has encouraged the development of indicators and monitoring tools concerning sustainability. 
There have been several efforts in developing such  dashboards while considering its multi-dimensional nature. Dashboard for Sustainability~\cite{Scipioni2009} is one such tool to represent sustainability using indices. The tool aggregates different indicators, taking into account the level of importance, and represents them as scores on a single scale, ranging from bad to good. This tool was used in Padua to monitor local sustainable development as per the action plan of the municipality. The study concludes that the dashboard's effectiveness lies in the definition of indicators and the relative weight assigned to them. In addition, indicators must be monitored over an extended period of time to perform a realistic evaluation of progress.
Sardain \etal~\cite{Sardain2016} present a sustainability dashboard to monitor national sustainability indicators in Panama, provided by a wide range of stakeholders. Their approach employs correlation coefficients to identify inter-indicator trade-offs. 

Three of the reviewed works present the process they use to establish their KPIs and dashboards~\cite{WOLFERT2022,Vannieuwenhuyze2020,Sardain2016}; two of them are reports of the ad hoc methods used~\cite{Vannieuwenhuyze2020,Sardain2016} and one with a generic process focusing on Challenge 1~\cite{WOLFERT2022}. None of the reviewed works formalized a generic process addressing Challenge 2. In our work, we propose a generic process for both challenges. 
Moreover, none of the reviewed works considered a federation where each part of the federation has a different focus, as we do in our work. Three considered only one city/country~\cite{Vannieuwenhuyze2020,Sardain2016,Scipioni2009}. The remaining two consider different measurement sites, but all had the same focus and used the exact same set of KPIs~\cite{WOLFERT2022,FREITAS2025}.   

\section{The \feedForfood project}
\label{sec:Project}

The \feedForfood project aims to empower vulnerable communities in three \european cities 
through urban agriculture \LLs, to pioneer a shift to sustainable and inclusive urban agriculture. \LLs, as defined by the European Network of Living Labs, are ``real-life test and experimentation environments that foster co-creation and open innovation among the main actors of the Quadruple Helix Model: Citizens, Government, Industry, and Academia''~\cite{ENoLL_definition}. The project is built on three pillars, which are: (i) the three \LLs, (ii) a Central Knowledge and Learning Hub (CKLH), and (iii) a transition strategy. 
The \LLs are where the urban gardening/farming and related activities are taking place. A wide range of data is collected in the \LLs, which is analyzed by the CKLH. The knowledge thus extracted is the basis of the transition strategy, which will ensure that the \LLs can continue their mission after the end of the \feedForfood 
project. 

\subsection{Vision}
\label{sec:Vision}

\begin{figure}
    \centering
    \includegraphics[width=\linewidth]{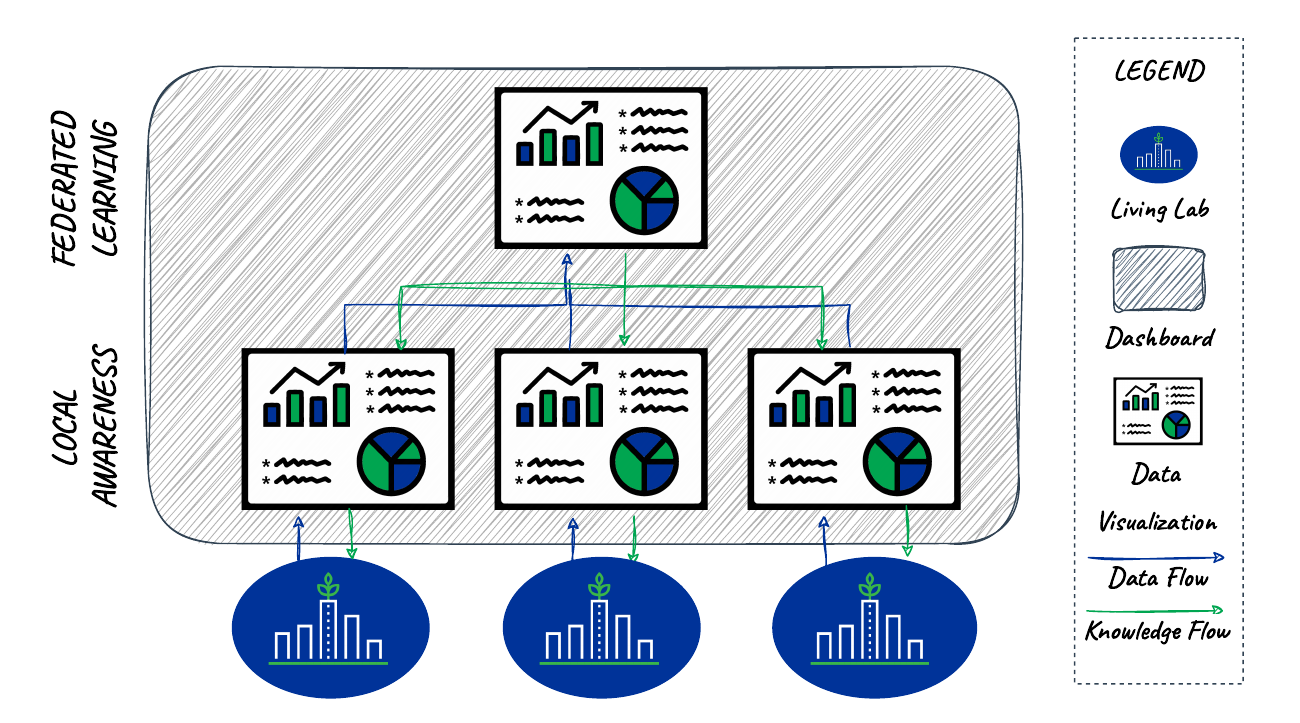}
    \caption{The vision of the \feedForfood project}
    \label{fig:Vision}
\end{figure}

Each \LL has its own challenges and ways of addressing them. The vision of the \feedForfood project, illustrated in Fig.~\ref{fig:Vision}, is to leverage the differences and similarities of the \LLs inside a federation. The federation consists of various \LLs and a common tool, which is called the \textit{\feedForfood dashboard}. This tool is able to receive, analyze, and visualize data. Each \LL sends data to the dashboard about its activities, production, and other relevant parameters. The dashboard provides each \LL with a display to monitor relevant KPIs, and extract knowledge about how to successfully and sustainably achieve the objective of the \LL. This is the local awareness level, where each \LL becomes aware about how it is doing with regards to its own objective(s). 
At a second level, the data from all the \LLs is analyzed in order to learn from the federation and to reuse knowledge and good practices between \LLs. The knowledge obtained from the federated learning level is communicated back to the \LLs through separate displays. As all tools developed during the project, the dashboard itself should be sustainable and inclusive, which leads to four challenges.  
First, the \LLs should clearly identify their goals and what indicators should be monitored in the dashboard in order to track the progress towards the goal. Second, the data that the \LLs collect and upload to the dashboard should consist of all, and only, the relevant data. Therefore, the set of relevant data 
should be clearly defined and the corresponding measures identified. Third, the visualizations offered by each dashboard need to be understandable and actionable for \LLs so that \LLs can identify and decide the necessary actions to progress towards their specific goals. Thus, visualizations should be able to create awareness of \LLs. 
Finally, it is challenging to find ways to integrate the knowledge obtained through the dashboard within the federation so that \LLs can avoid pitfalls and reuse the good practices developed or identified
in other \LLs.

The \feedForfood federation of \LLs differs from usual user groups that share a common dashboard. First, the envisioned dashboard users are both enterprises and non-profit organizations or municipalities, which requires tailoring to both types of users~\cite{maheshwari_dashboards_2014}. 
Secondly, the different \LLs are structured in different ways and have different maturity statuses. Therefore, they are not several entirely comparable instances and do not have the same capacities for data collection.  
Third, although the \LLs follow the same overarching goal (sustainable and inclusive urban food systems), they focus on different target groups and ways to achieve the goal. 
This unusual combination requires a specific process for the establishment of the dashboard. 

\subsection{Specifics of each \LL}

The specific context of each \LL is further detailed in~\cite{Feed4FoodDeliverableD1-1} and summarized below.

In \bucharest, \Romania, the \LL focuses on the inclusion of migrants, the elderly, poor members of society, and the \Roma population. The integration challenges vary, with \Moldovans benefiting from cultural and linguistic similarities, while \Arab immigrants face more significant barriers. The elderly population is growing, and this demographic shift presents significant challenges. as the elderly often experience a decline in quality of life due to reduced mobility, reduced income, increased medical conditions, and psychological issues such as loneliness. Elderly farmers use land individually rather than in associations and are devoid of technological and material resources. \Roma populations suffer from discrimination and exclusion, which exacerbates poverty and food insecurity. The \LL in \bucharest aims to revitalise the peri-urban farms that have been neglected after \FallCommunism and turn these farms into places where the target groups can feel included and appreciated, while also increasing their food and nutrition security, as well as that of the wider \bucharest population.

In \drama, \Greece, the \LL focuses on the inclusion of women, dropped-out youth, and migrants. Here, 250 Social Vegetable Gardens  are revitalized and established, jointly employing members of the target groups, prioritizing women with a large number of dependent family members and unemployed members of society. Given the very high unemployment among young people, the gardens also aim to reach dropped-out youth, increasing their skills and confidence to allow a return to the regular labor market or the start of small-scale enterprises. In addition, the gardens will produce fresh fruits and vegetables, increasing food and nutrition security for poorer segments of the \drama population. A special challenge for the \LL in \drama is the selection of appropriate crops and crop varieties that are suitable for the current and expected future climate. The use will be made of traditional, sometimes ``forgotten'' varieties, as they in many cases are more resilient and high in nutrients.

In \strovolos, \Cyprus, the \LL comprises one location, and has close collaboration with an existing NGO, the \SteliosChristosIoannou foundation, which is already running an urban agriculture initiative focusing on people with mental disabilities and mental health problems. The target groups of the \LL are the elderly, poorer populations, migrants, mentally challenged people, and unemployed citizens. Many seniors seek opportunities to remain active and engaged in their communities. Urban farming allows them to participate in physical activities, contribute their life experiences, and enjoy the therapeutic benefits of gardening. The poor population benefits significantly from urban farming by gaining access to affordable, fresh produce and opportunities to reduce household food expenses. Migrants (\Africans, \Syrians, \Asians) are a significant group, often facing challenges such as language barriers, limited access to resources, and social exclusion. By involving them in the \LL, urban farming provides a platform for integration, skill development, and a sense of belonging within the local community.  Urban farming offers mentally challenged people a safe, inclusive environment where they can learn new skills, connect with others, and gain a sense of achievement. Unemployed people are supported through the creation of job opportunities within urban farming, such as production, maintenance, and distribution roles. The \LL provides training and pathways to employment, enabling participants to gain valuable skills and income-generating prospects.
\subsection{Actors and Roles}

The \feedForfood project puts together a diversity of actors. They  include persons affiliated with one of the \textit{project partners}, referred to for conciseness as project partners. They can be researchers, municipal or association workers, \etc. The authors of this work are researchers affiliated with the university responsible for the CKLH. In addition to these, the \LLs will be used by different persons, called the \textit{\LL participants}. These may belong to a vulnerable group, in which case they are part of one of the \textit{\LL target group(s)}. All the persons connected to the \LLs, either as participants or through project partners, as called \textit{stakeholders}. 

During the project, the stakeholders can fulfill different roles. The roles that are relevant to this work are the following. The authors of this work fulfill the role of \textit{dashboard designers}.  Selected project partners attended the different workshops organized as \textit{workshop participants}.
Once the dashboard is in use, some stakeholders will be \textit{data collectors}, collecting the data in the \LL or from other data sources. This data will then be input in the dashboard by \textit{data uploaders}. Some persons may be both but not necessarily. All stakeholders interacting with the dashboard are \textit{dashboard users}.     
\section{The Dashboard Establishment Process} 
\label{sec:ProcessDescription}
\begin{figure}[!htb]
    \centering
    \includegraphics[width=\linewidth]{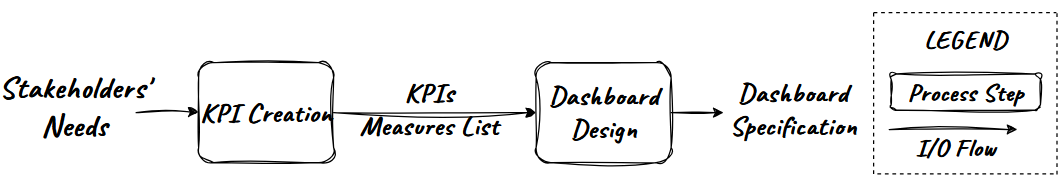}
    \caption{The process proposed in this work}
    \label{fig:process}
\end{figure}

To support the \LLs in achieving their objectives, the CKLH's mission is to establish a dashboard.  
In this work, a reusable process is proposed to address the two challenges described in Section \ref{sec:Introduction}, for a federation of different users. This two-step process is illustrated in Fig.~\ref{fig:process}. 
It is the first part of a longer process to achieve the vision presented in Section \ref{sec:Vision}.

The process takes the \textit{Stakeholders' Needs} as input. These needs may be clearly defined or still somewhat vague. The first step of the process will contribute to refining them. 
This first step is the \textit{KPI Creation} process. It addresses the challenge of collecting the relevant data. This process is explained in more details and exemplified for the \feedForfood project in Section \ref{sec:ViewKPI}.
The second step is the \textit{Dashboard Design} process. It takes as an input the \textit{KPIs} developed in the first step along with the \textit{Measures List}. It addresses the challenge of providing the right information to the dashboard user. This process is explained in more details and exemplified for the \feedForfood project in Section \ref{sec:ViewArchitecture}.
The outcome of this process is a \textit{Dashboard Specification} document specifying what should be implemented in the dashboard to start using it. After implementing it, a dashboard Minimum Viable Product (MVP) is obtained. 

\section{The KPI Creation Process}
\label{sec:ViewKPI}

Fatima \etal~\cite{Fatima2024KPIs} build on Parmenter's work~\cite{parmenter2015} to develop a \textit{KPI framework} that facilitates the creation of sound KPIs. According to the \textit{KPI framework}, the creation of KPIs starts with defining a long-term and high-level goal. This goal is broken down into Critical Success Factors (CSFs).  CSFs are critical areas where success leads to achieving the desired goal~\cite{Bullen1981}. To monitor the level of this success, each CSF is translated into a KPI.  A KPI is a function of one or more metrics, calculated using atomic measures. These metrics and measures can be reused across multiple KPIs if needed. Each KPI has a target value and an associated action. The KPI must be monitored frequently and continuously to observe the achievement of the relevant targets. If the target is not achieved, the action(s) must be triggered to reach the target.
\subsection{Process Description} 
\label{sec:KPI_StudyDesign}
\begin{figure}[!htp]
    \centering
    \includegraphics[width=\linewidth]{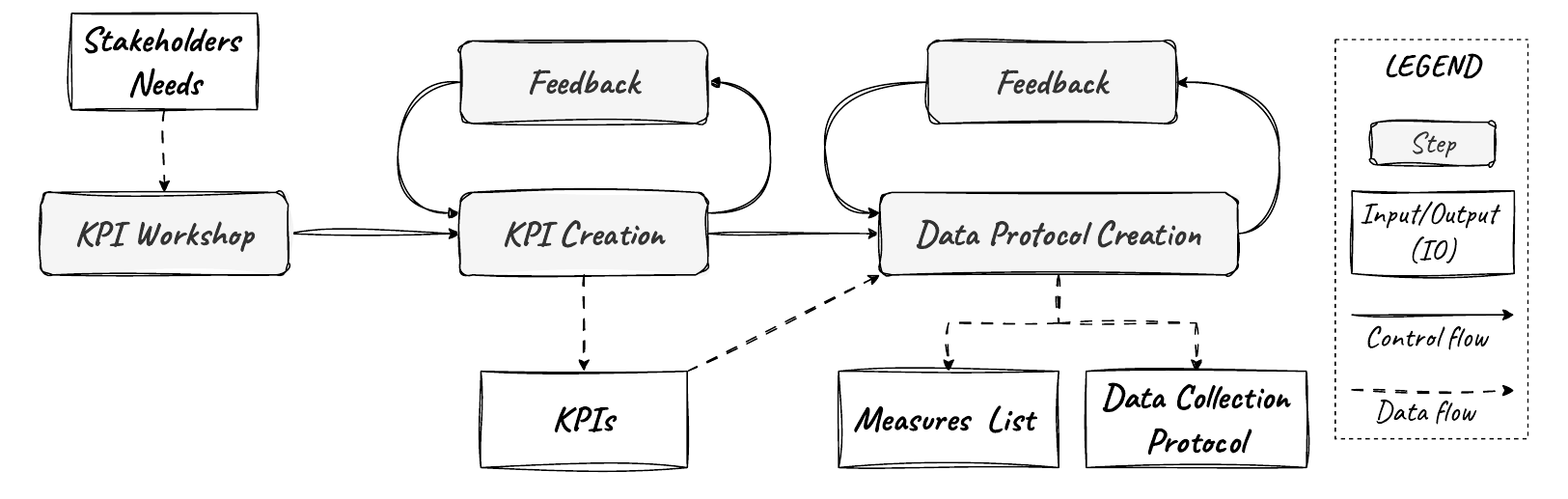}
    \caption{KPI Creation process}%
    \label{fig:KPI-steps}%
\end{figure}

The steps of the KPI Creation process (illustrated in Fig.~\ref{fig:KPI-steps}) are elaborated as follows.

\textbf{(i) KPI Workshop.}~The process starts with a workshop with the stakeholders where the dashboard designers act as facilitators. The aim of this workshop is to introduce the stakeholders to the KPI framework and start the KPI creation with the the elaboration of draft KPIs according to the needs of the stakeholders. The workshop can therefore be organized as follow: 1) An overview of the KPI framework is given, 2) each stakeholder group works on drafting a KPI, 3) the facilitators give feedback on the drafts and facilitate a short brainstorming session to refine the KPIs.

\textbf{(ii) KPI Creation.}~Then, the stakeholders further work on creating their KPIs independently. The dashboard designers provide  iterative feedback over several sessions, for possible improvements. During this feedback, the dashboard designers support the stakeholders in re-aligning the KPIs with their goals, identifying the measures for data collection, and aligning the functions for quantifying metrics and KPIs between the stakeholder groups.

\textbf{(iii) Data Protocol Creation.}~ Once the KPIs are refined and aligned, the list of all the different measures needed can be created. As a long-term aim is to be able to compare the different stakeholders group through the KPIs, there needs to be an agreement among the groups about how and how often the data collection should take place. Also, as it may be impractical for all groups to collect all the measures, it should be decided which subset of the measures is to be collected by all the groups, even if these are not directly connected to the KPIs they defined. The dashboard designers therefore propose a Measures List, which contains the names and frequency of the measures to be collected, separating the \textit{common} measures from the \textit{specific} ones. The groups give iterative feedback about this list. This list, together with the KPIs and additional concrete instructions about how to collect the data are documented into the Data Collection Protocol. The aim for this document is to act as a guide for helping the groups for the data collection. It should also act as a reference to limit the discrepancies in the ways the data is collected in order to obtain comparable data between the groups.

\subsection{KPIs monitoring the \LLs}
The process described in Section \ref{sec:KPI_StudyDesign} was applied for the \feedForfood project. Each \LL was a stakeholder group. 
The \LLs operate under the high-level goal of \textit{`sustainability'}. However, sustainability is a multi-dimensional concept. Hence, each \LL picked a scoped sustainability goal and defined the critical success factors, the KPI, its targets and actions along with the metrics and measures used to calculate it. In total, eleven KPIs were created by the \LLs and the CKLH, targeting three different sustainability dimensions\cite{OnlineMaterial}. The KPIs are summarized in Table \ref{tab:KPI_summary}. In this section, we present three of these KPIs (highlighted in \textbf{bold} in the table), one per sustainability dimension. 
\begin{table}[!htbp]
    \centering
    \caption{KPIs defined in the dashboard inception phase}
    \label{tab:KPI_summary}
    \begin{tabular}{|p{0.5cm}|p{5cm}|p{3.4cm}|l|}
    \hline
          \textbf{ID}& \textbf{Name} & \textbf{Sustainability Dimension} & \textbf{Created by}  \\
          \hline
         \textbf{KA1}& \textbf{Economic viability} & Economic & CKLH\\
         \hline
         KB1& Local vegetable access & Social & LL \bucharest\\
         \hline
         KB2& Pesticide use & Environmental & LL \bucharest\\
         \hline
         KB3& Agricultural employment& Social & LL \bucharest\\
         \hline
         \textbf{KC1}& \textbf{Effective training} & Social & All LLs\\
         \hline
         KD1& Target groups use& Social & LL \drama\\
         \hline
         KD2& Native species cultivation& Environmental& LL \drama\\
         \hline
         \textbf{KS1}& \textbf{Soil health} & Environmental& LL \strovolos\\
         \hline
         KS2& Water use efficiency & Environmental& LL \strovolos\\
         \hline
         KS3& Local and nutritious food production & Environmental& LL \strovolos\\
         \hline
         KS4& Target groups use & Social & LL \strovolos\\
         \hline
    \end{tabular}
\end{table}

\textbf{Soil Health KPI}

\begin{figure}[!hb]
    \centering
    \includegraphics[width=\linewidth]{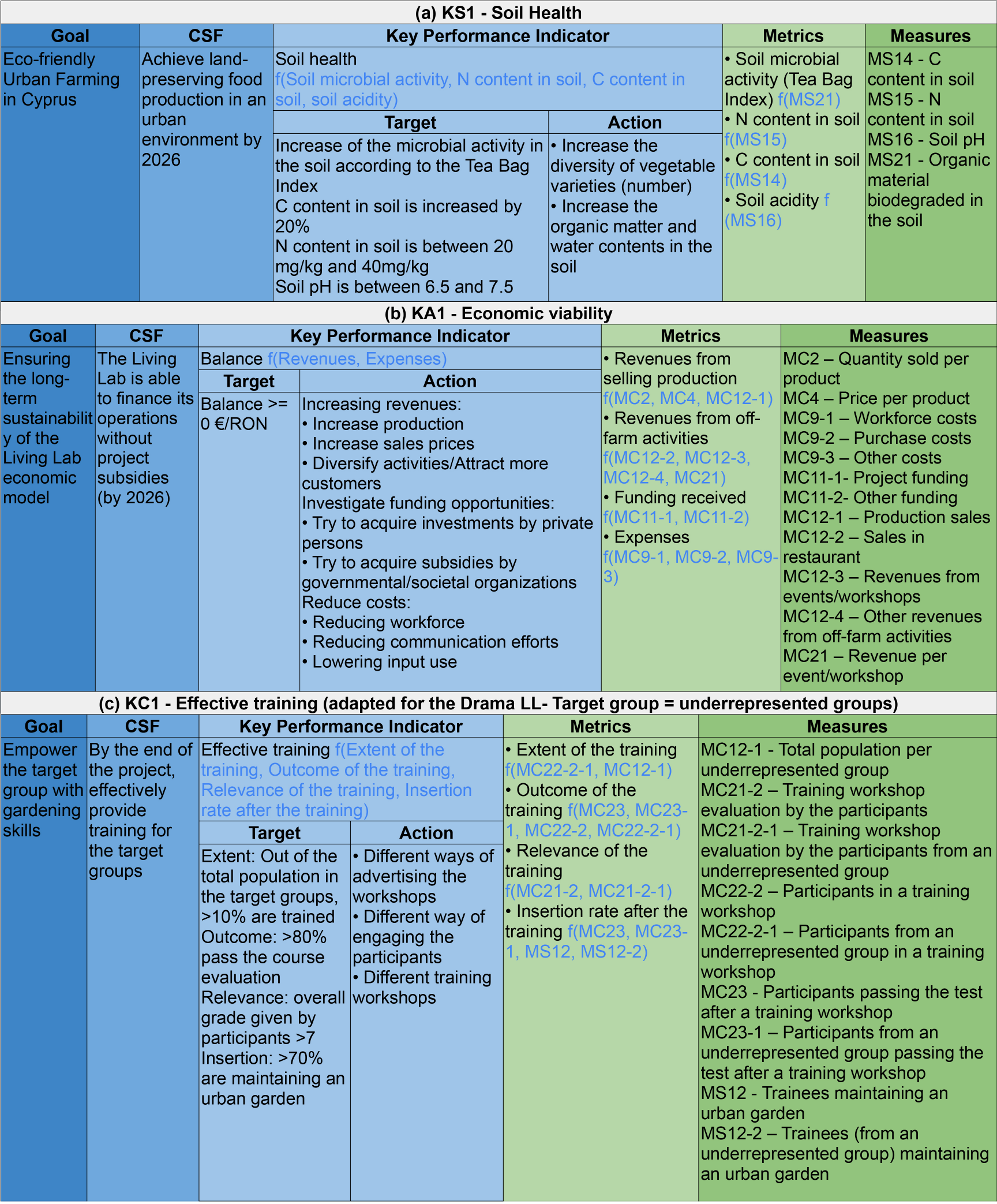}
    \caption{KPIs monitoring the LLs}
    \label{fig:kpis}%
\end{figure}

This KPI is designed to monitor soil health for environmental sustainability. Fig.~\ref{fig:kpis}(a) shows the key elements of this KPI. To ensure eco-friendly urban farming, the \LL must achieve land preservation during food production. To this aim, stakeholders develop a soil health KPI which is a function of four metrics; microbial activity in the soil, Nitrogen (N) retained in the soil, Carbon (C) retained in the soil, and soil acidity. These metrics are calculated using a list of measures (see Fig.~\ref{fig:kpis}(a)). The Soil Health KPI follows a conjunctive target, which means that success is only achieved if all four metrics meet or exceed their defined thresholds. Each metric must independently reach its required value for the KPI to be considered successful. If any metric falls below its threshold, the overall target is not met and actions must be taken such as increasing organic matter in the soil.  

\textbf{Economic Viability KPI}
\label{sec:EconomicKPI}

This KPI is designed to monitor the economic sustainability of the \LLs. Fig.~\ref{fig:kpis}(b) shows the key elements of this KPI. To ensure long-term sustainability (the transition strategy), it is crucial for each \LL to finance its operations without relying on project subsidies. To this aim the CKLH uses a \textit{Balance} function which must always have a positive outcome to meet the target. This balance is calculated as the difference between the generated revenue and expenses.  This KPI has a single value based target. In Fig.~\ref{fig:kpis}(b), the \textit{Metrics} column shows the data needed to calculate the KPI. The \textit{measures} column shows the data collected at the source by the \LL, which is used to calculate the metrics. If the target is met, \textit{actions} must be taken, such as increasing revenues, investigating funding opportunities, and reducing costs. 
Although each \LL may operate under a unique economic model, this KPI serves as a common way to evaluate financial success. However, specific actions taken may vary based on the lab-specific economic model. 

\textbf{Effective Training KPI}
\label{sec:TrainingKPI}

This KPI is designed to monitor the effectiveness of training activities in the \LLs. Fig.~\ref{fig:kpis}(c) shows the elements of this KPI. This KPI helps monitor the social sustainability goal \ie empowering target groups with gardening skills. The achievement of this goal is indicated by the effectiveness of training conducted throughout the whole length of the project. The KPI for effective training is a function of four metrics; extent of training, outcome of training, relevance of training, and insertion rate of trained individuals in the employment after training. Again, this KPI also has a conjunctive target subject to the achievement of a specific target value per metric. These metrics are calculated using the measures directly measured in the \LL. If the KPI fails to meet its targets, actions are planned such as better training advertisement, participant engagement, and introducing different types of workshops. 

\section{The Dashboard Design Process} 
\label{sec:ViewArchitecture}

\subsection{Process Description} \label{sec:desproc}

\begin{figure}[!htbp]
    \centering
    \includegraphics[width=\linewidth]{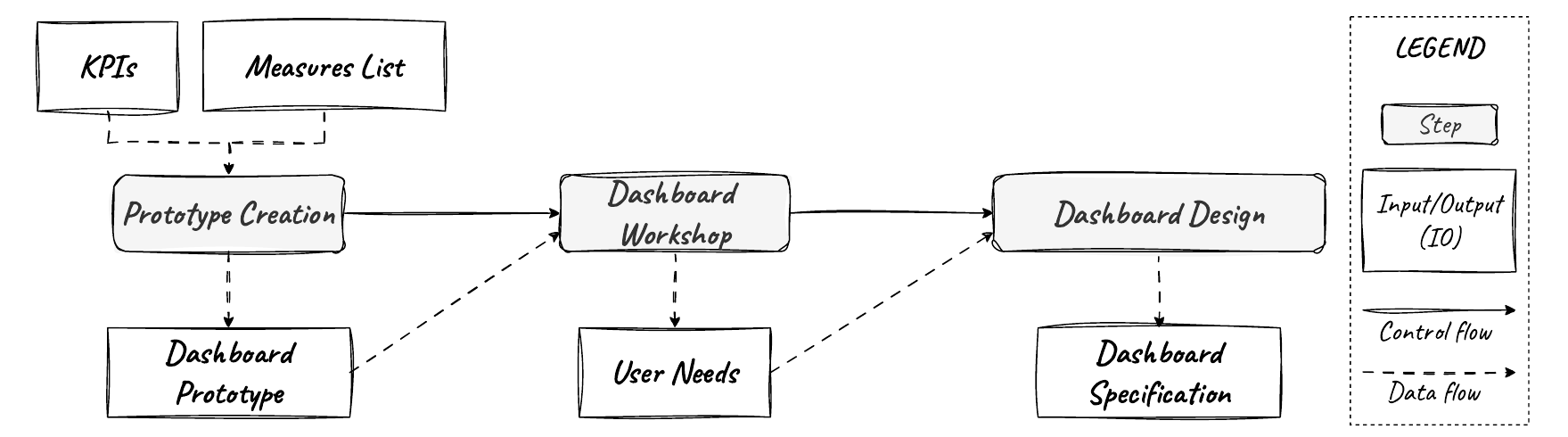}
    \caption{Dashboard Design process}
    \label{fig:DashBoardDesignProcess}%
\end{figure}

The steps of the Dashboard Design process (shown in Fig.~\ref{fig:DashBoardDesignProcess}) are elaborated as follows. The process can be repeated several times to focus on different parts of the design each time. 

\textbf{(i) Prototype Creation. } The first step is to create a prototype that can be shown to the stakeholders in order to gather their feedback on a visually concrete artifact. 
The prototype does not have to be actually working but should demonstrate an example workflow of using (a part of) the dashboard. A prototype can focus on a specific part of the dashboard. 

\textbf{(ii) Dashboard Workshop. } During a Dashboard Workshop, the prototype created in the previous steps is presented to relevant stakeholders. Feedback is gathered to identify the needs of the (different) dashboard users. 

\textbf{(iii) Dashboard Design. } Based on the user needs identified, the dashboard designers specify how the dashboard should be built to meet the user needs. The artifact created in this step, the dashboard specification, is then communicated to the implementation team. 

In the \feedForfood project, two dashboard design process iterations were completed. The first one focused on getting feedback on the overall structure of the dashboard, and on the data gathering part. The second one focused on the data visualization part.

\subsection{User needs}
For each iteration, a demonstration of a dashboard prototype was performed during a workshop, with mostly the same participants as for the KPI workshop. The workshops participants mainly gave oral comments after the demonstration. For the first iteration, they additionally filled a questionnaire containing three guided scenarios to gather information about the characteristics of the envisioned data uploaders (as these are a wider group than the workshop participants).

During the first iteration, the high-level architecture of the dashboard, \ie through a website, with one interface for gathering data and a second (connected) interface for visualizing the data was given positive feedback by the participants. No alternative was suggested. 

For uploading data, the three options showcased in the first iteration prototype were: 1) input form for an individual measure, 2) input form for a collection of related measures (called a \textit{report}), for example, about production yield and rainwater harvested during a given day, and 3) file upload. The participants found these three options relevant. They mentioned that they regularly collect data through forms using third-party tools (\eg Google Forms) so a way to upload such form data into the dashboard would be beneficial.

The outcome of the first iteration questionnaire was that three main categories of users are expected to upload data: researchers, municipality workers, and \LLs participants. They will have different levels of ease using digital tools, different language preferences, and will use different types of devices to connect to the dashboard (\eg smartphones or laptops). The data is likely to be first stored on paper or in a file (\eg Excel) before being input into the dashboard portal.

Regarding the visualization of the data, the one presented in the first iteration prototype 
was very customizable by the user, with the possibility to create their own visualizations. The workshop participants reacted that this is useful for researchers but likely too overwhelming for other stakeholders, for whom a more standard infographic-like visualization is better suited. This is in line with previous work~\cite{pluto-kossakowska_dashboard_2022}. In the second iteration prototype, two non-customizable categories of graphs were presented. First, graphs that can be used for monitoring each of the KPIs and second, graphs monitoring the raw data that a specific data uploader entered. The aim of the first category is to support awareness creation and decision support at the \LL level. The aim of the second category is to provide a data uploader with a way to monitor its own data, as an incentive for participating in the data collection. The workshop participants were very positive to the second set of graphs and they expressed that it would increase the benefits of the tool considerably. 

In a nutshell, the following central needs should be driving the further development of the dashboard. According to the project vision, the dashboard itself should be sustainable. At a high-level, a website should be developed with one data gathering part and one data visualization part.  More specifically, for data gathering:
\begin{itemize}
    \item The interface should be easy to use, including for user with disabilities
    \item It should handle different types of data sources, including forms and tabular files
    \item The interface should be accessible both on mobile and stationary devices.
\end{itemize}
And for data visualization:
\begin{itemize}
    \item Different visualizations and different levels customization should be proposed for different types of users. 
\end{itemize}

\section{Insights and Challenges}
\label{sec:Discussion}

\subsection{Insights}

We observed that the generic nature of the KPI framework~\cite{Fatima2024KPIs} helped in its adoption in the context of \LLs. In the past, this framework was only used to develop technical KPIs. 
This work is therefore the first one to apply this method with public sector organizations and looking at KPIs that are not solely related to a technological product. The study presented here demonstrates that the KPI framework indeed is applicable to a wider range of domains than the ones initially covered.

Trade-offs among KPIs can 
manifest over time. For example, for KS1 (see Fig.~\ref{fig:kpis}(a)), there is a trade-off between maintaining soil health and maximizing crop yield. Replacing chemical fertilizers with organic alternatives can improve soil health, leading to nutrient crops. However, this may result in a decrease in yield. With continuous monitoring over time, \LLs can observe the impact of their actions and take decisions that balance these trade-offs. 

The common goal of the \LLs in the \feedForfood project is to achieve inclusive, healthy, and sustainable urban food environments. At the KPI workshop, the workshop participants were given the instruction to create KPIs related to this broad goal. It is interesting to note that that the KPIs they chose to create reflect the specific characteristics of each \LLs. This confirms the ability of the methods to tailor the indicators to their specific situation. It can be noted that none of the KPIs proposed by the \LLs was targeting the economic dimension. Yet, economic viability is essential for the continuation of the \LLs after the project, and a KPI about this was therefore added at the initiative of the CKLH (KA1, see Fig.~\ref{fig:kpis}(b)).

The \LLs defined their KPIs separately, although they had access to the ones already defined by other \LLs in order to compare or get inspiration. As a result, two KPIs (KD1 and KS4) are similar, but not identical. 

When revising the full set of KPIs, we found that \textit{training} was present in at least one KPI for each \LL and therefore decided to extract it as a separate KPI (KC1, see Fig.~\ref{fig:kpis}(c)). 
We therefore have two kinds of common KPIs in this study: one defined externally to the \LLs (KA1), and one internally (KC1). For the first case, the KPI definition is the same for all \LLs, with a common list of measures. In the second case, the KPI definition is declined with one variant per initially defined KPIs as, even if the overall is the same, the actual measures and targets will differ for each \LL. This is possible to achieve thanks to the flexibility offered by the KPI framework which can be adapted to such situations. 

During the planning phase, an initial list of measures to be collected by the \LLs had been envisioned. Using the proposed process with the KPI framework allowed us to realize that many different measures needed to be added for evaluating the KPIs.  
Hence, using the KPI framework is likely to save time and effort by enabling a focused data collection from the start and by reducing the risk of having to change the collected data later on in the project.  

During the KPI refinement process, two pitfalls were observed. The first one is to define a KPI with a too broad scope. This manifests with a list of measures that cannot be measured within the \LLs or with targets that the \LL has none or limited influence on. 
The second pitfall is to include too many different dimensions or sub-indicators in the same KPI. Difficulties to find a concise name for the KPI that covers all the dimensions should be taken as a hint to consider whether the indicators presented in this KPIs are better suited to be presented in separate KPIs. For example, KS1, KS2 and KS3 were initially gathered in the same KPI with an initial name that was too broad to be suitable. When refining the name, the impossibility to find a short name covering all the parts lead us to split the content into three KPIs.

\subsection{Challenges and Way Forward}

As described before, the \LLs created their KPIs separately. Hence, their KPIs include different measures. To prepare for learning within the federation, it is however needed that the \LLs collect a common set of measures, to allow comparisons. This is challenging in two ways. First, it means that the \LLs have to put effort in measuring data that they may consider as useless for them (since it is not used in ``their'' KPIs). As the \LLs have limited resources, there is an initial reluctance to do so which has to be overcome. Second, even when there is an agreement to collect the same measure, it needs to be ensured that the data will indeed be collected in the same (or a similar enough) way. As the measuring environments are very different (and the CKLH has no direct access to them), it is a challenging task to provide appropriate and detailed enough data collection instructions through \eg the data collection protocol. 

At the Dashboard Design workshop, the participants were in agreement that the dashboard should be ``easy and straightforward to use''. Understanding what this concretely mean is challenging in the \feedForfood context for several reasons. To start with, who the users of the dashboard will be was still an open question during the inception phase (i.e. the phase described in this work). As the \LLs have not started their activities yet, we had to work on assumption on who the users will be, with no possibility to ask for their feedback on the prototype.   
Then, it has to be ensured that user feedback can be collected when the dashboard starts to be used. This is needed to improve the dashboard, to ensure that the dashboard implementation is not impeding the data collection, awareness creation or learning instead of supporting it, and that it keeps the users motivated to participate in the data collection process. Achieving this through indirect contacts is challenging.    

As a \feedForfood tool, the dashboard should also be sustainable and inclusive. Therefore, in addition to the KPIs already defined for assessing the performance of the \LLs, KPIs will be developed for the dashboard itself. They will measure how well the dashboard objectives are currently achieved by the MVP implemented based on the output of the proposed process and how the later improvements of the MVP will impact these KPIs. 
These new KPIs will address the technical sustainability dimension~\cite{Lago_framing}, which is currently missing.

In the dashboard inception phase, we set the stage with concepts and tools which we believe are going to enable decision support and awareness creation. Looking forward to the next dashboard phase, we have to evaluate to which extent these concepts and tools are actually fulfilling their objective. How to perform this assessment is a challenging task. 

The steps performed so far are mainly targeting the local level, where each \LLs is considered a different node in the federation (the middle level of Fig.~\ref{fig:Vision}). However, the vision  of the project and the focus of phase 3 is to explore how learning can be achieved within the federation (the top level of Fig.~\ref{fig:Vision}). Which methods have to be implemented in the data analysis module to support this goal and whether the current KPIs and data collection list are comprehensive enough to support federated learning are two challenging tasks which are part of the next steps of the work.

\section{Conclusion}
\label{sec:Conclusion}

In this work, we propose a process that enables establishing a dashboard to support a sustainable and inclusive urban agriculture federation. We solve the challenges at hand with two sub-processes that (i) identify what data should be collected leveraging the KPI framework, and (ii) create a dashboard prototype to determine user needs for visualization. We exemplify these two sub-processes in the case of the \feedForfood project. 

As future work, 
we will explore how to evaluate the created awareness and how to achieve federated learning. The dashboard implementation will be further developed to support this, as well as other relevant needs from the stakeholders. In addition, KPIs for the dashboard implementation will be created. 

 \printbibliography

\end{document}